\documentclass[aps,prl,twocolumn,superscriptaddress,showpacs]{revtex4-1}
\usepackage{graphicx}
\usepackage{amsmath}
\usepackage{amssymb}
\usepackage{bm}
\usepackage{pifont}
\usepackage{color}
\usepackage[FIGTOPCAP]{subfigure}
\usepackage[latin1]{inputenc}

\begin{document}

\title{Body size affects the strength of social interactions and spatial organisation of a schooling fish ({\it Pseudomugil signifer})}
\author{Maksym Romenskyy}
\affiliation{Department of Mathematics, Uppsala University, Box 480, Uppsala 75106, Sweden}
\author{James E. Herbert-Read}
\affiliation{Department of Mathematics, Uppsala University, Box 480, Uppsala 75106, Sweden}
\affiliation{Department of Zoology, Stockholm University, Stockholm 10691, Sweden}
\author{Ashley J. W. Ward}
\affiliation{School of Biological Sciences, University of Sydney, Sydney, New South Wales, Australia}
\author{David J. T. Sumpter}
\affiliation{Department of Mathematics, Uppsala University, Box 480, Uppsala 75106, Sweden}
\date{\today}

\begin{abstract}
While a rich variety of self-propelled particle models propose to explain the collective motion of fish and other animals, rigorous statistical comparison between models and data remains a challenge. Plausible models should be flexible enough to capture changes in the collective behaviour of animal groups at their different developmental stages and group sizes. Here we analyse the statistical properties of schooling fish ({\it Pseudomugil signifer}) through a combination of experiments and simulations. We make novel use of a Boltzmann inversion method, usually applied in molecular dynamics, to identify the effective potential of the mean force of fish interactions. Specifically, we show that larger fish have a larger repulsion zone, but stronger attraction, resulting in greater alignment in their collective motion. We model the collective dynamics of schools using a self-propelled particle model, modified to include varying particle speed and a local repulsion rule. We demonstrate that the statistical properties of the fish schools are reproduced by our model, thereby capturing a number of features of the behaviour and development of schooling fish.
\end{abstract}


\maketitle

\section{Introduction}
In sufficiently large collective systems, the behaviour of an individual can be dominated by the generic statistical effects of many individuals interacting, rather than its own behaviour \cite{Romensky20150015}. Much of the progress in understanding collective motion of animal groups has involved applying ideas borrowed from the statistical physics of materials like magnets or fluids  \cite{vicsek.t:2012,herbert-read.j:2011,herbert2015initiation,bialek.w:2011,toner.j:2005}. For example, changes in group densities produce phase transitions at critical group sizes \cite{buhl.j:2006,calovi.ds:2014}. More complex collective states, such as swarm, mills and polarised groups depend on the density of a group and the noise within the system \cite{tunstrom.k:2013}. Recent studies of starlings and midges have looked at spatial velocity fluctuations \cite{attanasi.a:2014,cavagna.a:2010}, long range correlations \cite{bialek.w:2014}, and diffusive \cite{cavagna.a:2013} and entropic characteristics of flocks \cite{cavagna.a:2014}. Other experiments with artificial particles have looked for similarities and differences between self-organised living matter and thermal equilibrium systems \cite{narayan.v:2007}. These latter approaches gather statistical information about self-organising structures in order to parameterise models (see for example the maximum entropy approach \cite{cavagna.a1:2014,cavagna.a:2010}). However, none of these have explicitly solved the inverse problem of using the macro-level properties of animal groups to find out how the individuals within them interact.

This inference problem is essentially a statistical physics problem. The last few decades have seen a major increase in research at the interface of molecular dynamics and biophysics. In soft matter systems, estimating the potential energy of an interaction and the corresponding potentials is of particular importance, as the strength of intermolecular interactions determines the state of matter and many of its properties \cite{meckelke.m:2013}. At the same time, molecular interaction potentials are difficult to measure experimentally and hard to compute from first principles. An alternative approach, therefore, is to estimate them from experimentally determined structures of molecules. The interactions in these structures are usually strongly coupled and assemblies are typically driven by weak forces (e.g., hydrophobicity or entropy) \cite{moore.t:2014}. Therefore, estimation of these potentials requires application of sophisticated coarse-grained techniques such as reverse Monte-Carlo \cite{mcgreevy.r:1988}, inverse Monte Carlo \cite{lyubartsev.a:1995} or Iterative Boltzmann inversion \cite{soper.a:1996}. These methods adjust the force field iteratively, until the distribution functions of the reference system are reproduced as accurately as possible. In other cases, when the potentials are uncoupled or weakly coupled, a more straightforward direct Boltzmann inversion approach can be applied, which approximates the potential by the negative logarithm of the radial distribution function \cite{tschop.w:1998}. In collectively moving animal groups the interactions between members are usually assumed to be of hierarchical structure, with repulsion having highest priority at small distances \cite{couzin.id:2002}. Thus one can expect that the latter method can be also applied to animal self-organised systems, such as fish schools, to infer the interactions within these groups from experimental data.

Here we investigate the schooling behaviour of fish using Boltzmann inversion and related methods. Unlike molecules and physical particles, fish change their behaviour as they go through various developmental stages \cite{Hinz13022017}. For example, onset of schooling is only possible when the central nervous system of fish is sufficiently developed to support a high level of coordination of visual and mechanosensory information \cite{masuda.r:1998}. The developmental differences are usually also reflected in changes of the key characteristics of motion, including speed. Therefore, fish of different sizes can not be considered simply as particles of different physical size, since their behaviour changes with their size. We thus expect the statistical properties of the group, and of individuals, to change both with the density of fish and their developmental stage.

\section{Methods}

\subsection{Experimental details}

We used groups of 10 to 60 Pacific blue-eyes (\emph{Pseudomugil signifer}) with approximately three different body lengths (from hereafter referred to as fish sizes): $\sim$ 7.5 mm (small), $\sim$ 13 mm (medium) and $\sim$ 23 mm (large) (see the electronic supplementary material figure S1). Because body size is related to the age of a fish \cite{pusey.b:2004}, the three body lengths used in this study likely represent three distinct age classes. The largest fish (23 mm) constituted of sexually mature individuals, although we observed no sexual behaviour in the trials. The fish were confined into a large shallow circular arena (760 mm diameter) and filmed from above at high spatial and temporal resolution. The positions of fish were subsequently tracked using DIDSON tracking software \cite{handegard.n:2008}. On average 86\% of fish were identified and tracked in our experiments, which is a similar level of accuracy as compared to other studies that track large numbers of individuals \cite{tunstrom.k:2013}.

\subsection{Model}

\begin{figure}
\centering
\includegraphics[width=7.0cm,clip]{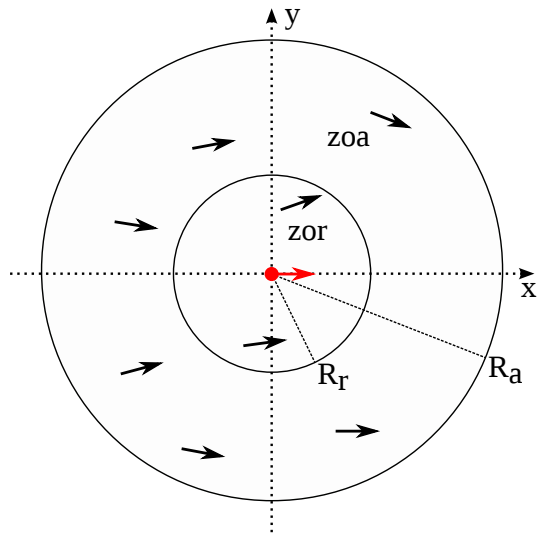}
\caption{The illustration of the interaction parameters in the SPP model. The particle shown in red turns away from the nearest neighbours within the zone of repulsion (zor) to avoid collisions and aligns itself with the neighbours within the zone of alignment (zoa).}
\label{fig:model}
\end{figure}

In our two-dimensional model, the fish are represented by $N$ point particles at number density $\rho$ and variable particle speed $v_i$. The system undergoes discrete-time dynamics with a time step $\Delta t$. The direction of motion of each particle (Fig. \ref{fig:model}) is affected by repulsive or alignment interactions with other particles located inside the zone of repulsion (zor) or zone of alignment (zoa), respectively. Time evolution therefore consists of two steps: velocity updating and streaming (position update). In the first computational step, position of each particle ($\mathbf{r}_i$) is compared to the location of the nearest neighbours. The repulsion rule has an absolute priority in the model and is modelled as a typical collision avoidance \cite{couzin.id:2002,romenskyy.m:2013} ${\hat{\mathbf{u}}(t)_{i}=- {\sum\limits^{n_r}_{j\neq i}\mathbf{r}_{ij}(t)}/{\left |\sum\limits^{n_r}_{j\neq i}\mathbf{r}_{ij}(t) \right |}}$ with $\mathbf{r}_{ij} =\mathbf{r}_{j}- \mathbf{r}_{i} $ and $n_r$ being a number of particles inside $zor$. The alignment rule similar to one used in the Vicsek model \cite{vicsek.t:1995} takes into account velocities of all particles located inside the zone of alignment ${\hat{\mathbf{u}}(t)_{i}={\sum\limits^{n_a}_{j=1}\mathbf{V}_j(t)}/{\left |\sum\limits^{n_a}_{j=1}\mathbf{V}_j(t) \right |}}$ with $n_a$ being a number of particles inside $zoa$. The velocities of particles are updated according to
\begin{equation}
\mathbf{V}_i(t)=v_i(t) \hat{\mathbf{u}}(t)_{i} R_1(\xi_i(t))R_2(\theta_i(t))
\tag{S7}\label{velocity}
\end{equation}
with $v_i(t)=v_0[\psi(t)]^\gamma$ defining the particle individual speed $v_i(t)$ based on the averaged local order $\psi(t)$ inside both behavioural zones \cite {li.w:2007,zhang.j:2009,mishra.s:2012,lu.s:2013}. $v_i(t)$ takes its maximal value $v_i(t)=v_0$ when velocities of particles inside zor and zoa are perfectly aligned $\psi(t)=1$ while absence of local order $\psi(t)=0$ results in $v_i(t)=0$. The exponent $\gamma$ controls the sharpness of the speed change. Note that for any $\gamma$ an isolated particle will move with maximal speed $v_0$. The misaligning noise is introduced through a random rotation $R_1(\xi_i(t))$ of the resulting particle velocity according to a Gaussian distribution $P(\xi_i(t))={e^{- \xi_i^2(t) / 2 \eta^2} / \sqrt{2 \xi_i(t)} \eta}$, where $\xi_i(t)$ is a random variable and $\eta$ is the noise strength.

Wall avoidance is modelled as a particle orientation adjustment through rotation $R_2(\theta_i(t))$ of the particle velocity with a time-dependent turning rate $\theta_i(t)=v_0 \phi_i(t)/d_i(t)$. $\phi_i(t)$ is the angle between the heading of a fish and normal to a time-dependent point of impact on the wall \cite{gautrais.j:2009,gautrais.j:2012}. $d_i(t)$ denotes a distance from particle $i$ to the impact point. Such construction of the rotating rate allows to achieve its strong damping at large distances from the wall and for smaller angles of approach of a collision point on the wall. At these conditions its influence on particle's motion is insignificant.

When the velocity update step is complete, the particle positions are updated by 
\begin{equation}
\mathbf{r}_i(t + \Delta t)=\mathbf{r}_i(t)+ \mathbf{V}_i(t)\Delta t
\label{motion}
\end{equation}

The model was parametrised based on the experimentally obtained data. The unit of length in our simulations is equivalent to the metric length used in the experiment. To set the unit of time we choose a particle speed $v_0=bv^{e}_0$, where $b$ is the behavioural reaction time \cite{domenici.p:1997} of fish ($b=0.05$ s) and $v^{e}_0$ is the average speed in experiment. The integration time step was set to $\Delta t =1$ for all simulations. The noise strength $\eta$ was fixed at 0.1 for all trials. The exponent $\gamma$ was set to 1 for in all simulations. Fish of different size are modelled by scaling the size of the alignment zone with a factor $k$ proportionally to experimentally measured differences in body lengths so that $k=1$, $k=1.73$ and $k=3.07$ correspond to small (7.5 mm), medium (13 mm) and large (23 mm) fish, respectively.

Total number of time steps in each run was $1 \times 10^6$. The statistics was collected in the steady state and each characteristic of motion was calculated by averaging over five independent runs. The radius of the arena was fixed at $R=380$ for all simulations. The initial conditions for fish positions and velocities were chosen at random from the uniform distribution.

\section{Results}

\begin{figure}
\centering
\includegraphics[width=\linewidth,clip]{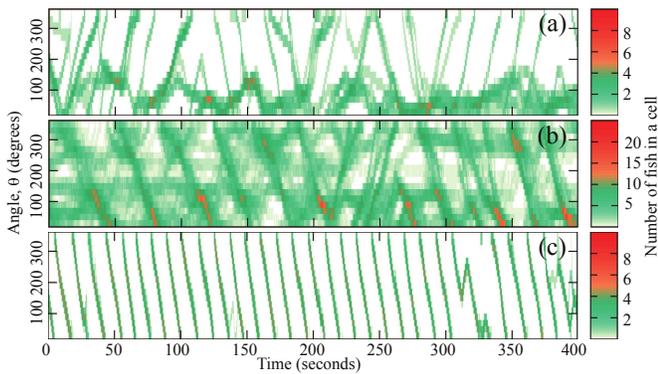}
\caption{Example time evolution of the spatial distribution in arena at $N=10$ small fish (a), $N=60$ small fish (b), and $N=10$ large fish (c). Each cell in plots (a)-(c) denotes a 20 degrees radial segment of the arena and 1 second of time. $\theta$ is the angle measured counterclockwise from the positive direction of the $x$-axis as defined by the camera position.}
\label{fig:spatial_dist}
\end{figure}

We first investigated the spatial distribution of fish in the arena. Small fish displayed limited collective motion. For example, 10 small fish tended to form a dispersed group, where most of the fish moved very little (figure \ref{fig:spatial_dist}(a)). Larger groups of 60 small fish showed slightly more collective motion, but not all fish moved in the same direction at the same time (figure \ref{fig:spatial_dist}(b)). In contrast, even small groups of large fish showed highly aligned collective motion (figure \ref{fig:spatial_dist}(c)).

\begin{figure}
\centering
\includegraphics[width=\linewidth,clip]{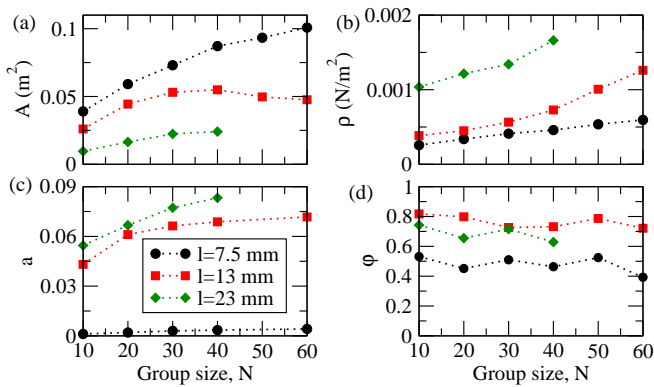}
\caption{Statistical properties for groups of fish with three different body lengths: $7.5$, $13$, and $23$ mm. (a) Area of a group $A$. (b) Number density $\rho$. (c) Packing fraction $a$. (d) Polar order parameter $\varphi$.}
\label{fig:A_rho_a}
\end{figure}

We calculated the average area, $A$, covered by the group using a convex hull algorithm (see the electronic supplementary material). The average value of $A$ is plotted for different group and fish sizes (figure \ref{fig:A_rho_a}(a)). For all three fish sizes, larger groups occupied a larger area. The density, $\rho=N/A$, also increased with the number of fish in the group (figure \ref{fig:A_rho_a}(b)), suggesting that the fish pack closer together in larger groups. Figure \ref{fig:A_rho_a}(a) indicates that groups of small fish occupied a larger area than the groups of medium-size  or big  fish. This finding supports the results of the spatial analysis (figure \ref{fig:spatial_dist}); small fish were more dispersed over the arena. As a result, groups consisting of small fish were less dense (figure \ref{fig:A_rho_a}(b)) than groups of larger fish.

To better quantify the spatial arrangement of groups, we measured their packing fraction $a$ (figure \ref{fig:A_rho_a}(c)). This is the ratio between the total body area of all fish in a group ($A_f=\sum^{N}_{i=1} A_i$) and the global area of a group ($A$): $a= A_f / A$, where $A_i$ is the body area of individual fish.
For all body sizes of fish, packing fraction increased with group size. Groups of smaller fish had the lowest packing fraction ranging between $0.001$ and $0.004$ for groups of $10$ and $60$ individuals respectively. In contrast, groups of medium-size and large fish had higher packing fractions of $a>0.043$ and $a>0.054$ respectively. The lowest packing fractions in groups of small fish are comparable to those observed in bird flocks \cite{ballerini.m:2008,cavagna.a:2008}, whilst the larger packing fractions approach those of some bacteria \cite{peruani.f:2012}.
In physical systems, small values of $a$ typically correspond to gases, while larger values ($a>0.4$) to liquids or crystals \cite{dinsmore.ad:1997}.  All packing fractions observed in our experiments, therefore, are comparable with an atomistic system in its gaseous state.  At the same time, the large differences between packing fractions for small and medium-size fish and for small and large fish, reaching one order of magnitude (t-test, $p<1\times10^{-6}$), suggest possible differences in other statistical characteristics of the system for varying fish size.

We next characterised ordering in our system using the polar order parameter \cite{vicsek.t:1995} 
\begin{equation}
\varphi=\left \langle  \frac{1}{N} \left |\sum_{i = 1 }^N \exp(\imath \theta_i) \right | \right \rangle,
\label{polar}
\end{equation}
where $\imath$ is the imaginary unit, $\theta_i$ is the direction of motion of individual fish, and $\langle \cdot \rangle$ denotes the time average. We should note that the polar order parameter is generally sensitive to the choice of confining geometry within which the agents move. For instance, for a small circular arena, the relationship between the radius of the arena and radial positions and speeds of the agents determines the maximum value of polarisation that can be reached in the system. In our experiments, we used a large arena with a radius exceeding the average group width for large fish by more than 3 times. Large fish also occupied on average a radial segment of only 30 degrees (see figure \ref{fig:spatial_dist}(c)). Thus we expect that the polar order parameter can provide meaningful results in characterisation of alignment in our experimental system. Values of $\varphi$ are plotted in figure~\ref{fig:A_rho_a}(d) as a function of group and fish size. Small fish (7.5mm) did not display much ordering for any group size with $\varphi \approx 0.4-0.55$, confirming previous observations (figures~\ref{fig:spatial_dist}(a),(b)).  Groups of medium-size fish  were the most ordered, with values of the order parameter ranging between $0.73$ and $0.84$ depending on group size. Large fish  displayed slightly lower order than medium-size fish. 

\begin{figure}
\centering
\includegraphics[width=\linewidth,clip]{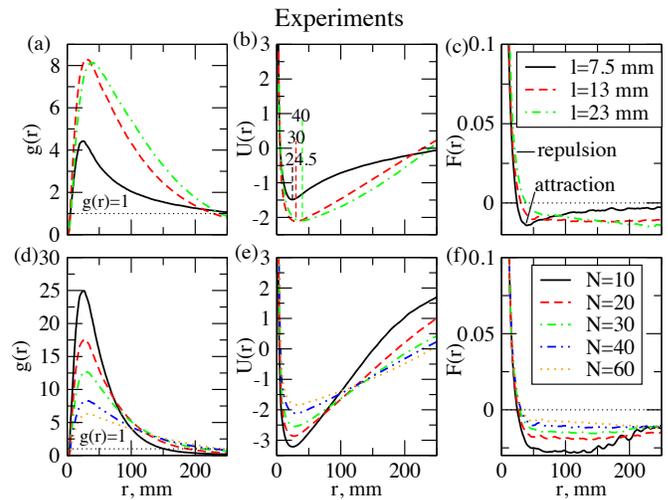}
\caption{Statistical properties of fish in experiments. (a) Pair distribution function $g(r)$ smoothed with 5-point moving average, (b) effective potential of the mean force of the interaction $U(r)$, and (c) mean force of the interaction $F(r)$ for small ($7.5$ mm) medium ($13$ mm), and large ($23$ mm) fish. (d) Pair distribution function $g(r)$, (e) effective potential of the mean force of the interaction $U(r)$, and (f) mean force of the interaction $F(r)$ for medium-sized fish ($13$ mm) in groups of different size. The top row contains plots for different body size and the bottom row contains plots for different group size. The top and bottom legends correspond to plots in the top and bottom rows, respectively.}
\label{fig:RDF_exp}
\end{figure}

We then investigated the nature of the interactions between the fish using statistical mechanics. We started by looking at the pair distribution function (PDF) \cite{romenskyy.m:2013} which allowed us to study how the local density around each fish varied with respect to the average density in the system. It is defined by
\begin{equation}
g(r)= \frac {1} {S(r)} \frac {1} {N (r)} \left \langle \sum_{i = 1 }^{N(r)} \sum_{j \neq i }^{N(r)}  \delta (r - |\mathbf{r}_{ij}|) \right \rangle,
\label{g_r}
\end{equation}
where $\delta$ is the Dirac delta function, $|\mathbf{r}_{ij}|$ is the distance between fish $i$ and $j$, $S(r)$ is the surface area of a shell, $N(r)$ is the number of fish inside a shell and $\langle \cdot \rangle$ stands for the time average (see the electronic supplementary material for details). A set of PDF-curves $g(r)$  for various fish sizes is presented in figure \ref{fig:RDF_exp}(a). Small fish tended to form aggregations with densities up to 4 times above the average density of the system and with a maximal half-radial width of more than 25 fish body lengths. For medium and large  fish, the maximum density in a cluster was 8 times larger than the average in the system. The size of the aggregation of medium and big fish was as large as 17 and 10 body lengths, respectively. Another notable difference between the three curves is the location of the local density peak. For small fish, the peak is at 24.5 mm, whereas for larger fish it is significantly shifted towards 30 - 40 mm. Figure \ref{fig:RDF_exp}(d) shows PDF plots for medium-sized fish at varying group size. For the smallest groups of 10 fish, the maximum density observed is 25 times above the system's average density. The peak value of $g(r)$ decreases with increasing group size (from 25 for groups of 10 fish down to 6.5 for the largest groups of 60 fish) whilst the position of the maximum remains unchanged at approximately 30 mm for all group sizes. The maximal half-radial width of the aggregation increases with group size from 145 mm for groups of 10 fish to 240 mm for groups of 60 fish. 

The differences in the pair distribution function suggest there is large variation in the underlying pair potential. In our system, in the absence of any external field, this potential is the effective potential of the mean force of the interactions between fish. Studies of active matter show that the resulting steady states of such systems often satisfy Boltzmann distribution \cite{bialek.w:2011}, even given that these active systems are essentially out of equilibrium. Here we apply the opposite route: we start with the assumption that the steady-state configurations observed in the experiment are drawn from the Boltzmann distribution
\begin{equation}
P(r)= Z^{-1} \exp{[ -\beta U(r) ]},
\label{Boltzmann}
\end{equation}
where $\beta=1/k_B T=1$ with a Boltzmann constant $k_B$ and the system's temperature denoted by $T$; $Z=\int \exp [-\beta U(r)]d r$ is the partition function and $U(r)$ is the effective potential of the interaction.  In this work, the choice of the inverse temperature $\beta=1$ is arbitrary as this term is not related to thermodynamic temperature in our system. Instead, it accounts for temperature-like fluctuations in the system which we assume to take the same value for all three size classes of fish considered here. The fact that we observe large differences in the stationary distributions of different sized fish and different group sizes is related to the changes in motion of individuals when the configuration of neighbours changes. Nevertheless, the average number of individuals within each shell of the pair distribution function remains constant over the whole duration of an experiment, defining a steady-state in our system.

\begin{figure}
\centering
\includegraphics[width=\linewidth,clip]{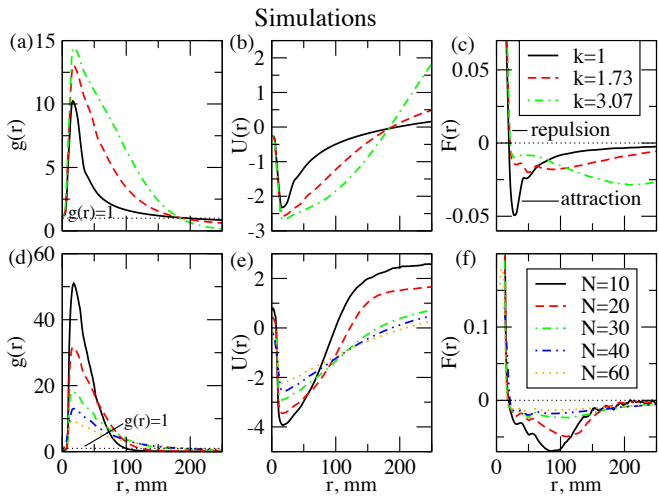}
\caption{Statistical properties of a model system. (a) Pair distribution function $g(r)$, (b) effective potential of the mean force of the interaction $U(r)$, and (c) mean force of the interaction $F(r)$ for small ($k=1$), medium ($k=1.73$), and large ($k=3.07$) simulated fish. (d) Pair distribution function $g(r)$, (e) effective potential of the mean force of the interaction $U(r)$, and (f) mean force of the interaction $F(r)$ for medium-sized fish ($13$ mm) in groups of different size. The top row contains plots for different body size and the bottom row contains plots for different group size. The top and bottom legends correspond to plots in the top and bottom rows, respectively.}
\label{fig:RDF_mod}
\end{figure}

To derive the effective potential of the mean force of the interactions we use the direct Boltzmann inversion \cite{tschop.w:1998}
\begin{equation}
U(r)= -k_B T \ln g(r).
\label{U_r}
\end{equation}
Figures \ref{fig:RDF_exp}(b),(e) present the effective potential energy profiles for different sized fish and different group sizes, respectively. Note that all the curves on both figures have practically the same slope for the decreasing part of the potential down to $U(r)=-0.55$. The minimum of these curves occurs at a greater distance for larger fish (figure \ref{fig:RDF_exp}(b)), indicating that larger fish have a larger repulsion zone. The increasing portions of the curves have similar slopes only for medium-sized and large fish (figure \ref{fig:RDF_exp}(b)). This suggests that fish of different body sizes and in groups of various size have similar repulsion strength (or collision avoidance potential) at short distances, but different attraction strengths towards neighbours. To get a conclusive picture of the variation of the interaction strength over the separation distance between the individuals, we calculate the mean force of the interaction $F(r)$:
\begin{equation}
F(r)=-\frac{d}{dr} U(r).
\label{E_r}
\end{equation}
In differently-sized fish (figure \ref{fig:RDF_exp}(c)), the repulsive force (positive and short portion of the $F(r)$ curves) is stronger than the attractive one (negative and long portion of the $F(r)$ curves). In other words, repulsion is independent of body size and spans a much shorter distance than attraction. Constant attraction force at large distances in our system arises because fish can perceive conspecifics over the entire arena. In other systems, a change of the attraction force over distance could be used for identifying topological interactions between individuals. Previous studies \cite{pearce} used specific channel confinement that mimics a geometrically frustrated anti-ferromagnet to distinguish between different types of interactions. Our approach is simpler and is not limited to specific experimental setups. For example, an abrupt change in the strength of attractive force would correspond to metric-type interactions.

To validate the fish interactions established from the experiments, we simulated the collective motion of the fish using a two-dimensional metric self-propelled particle model that accounts for variable fish speed and geometrical confinement (see Methods for model description and the electronic supplementary material for details on other motion statistics). Figures \ref{fig:RDF_mod}(a)-(c) present plots of PDF, effective interaction potential and mean force of the interaction for the three different sizes of simulated fish. Overall these qualitatively match the experimental data. Medium ($k=1.73$) and large ($k=3.07$) simulated fish form more dense groups than the small simulated fish ($k=1$) with a density of 12.5, 14 and 10 times above the average, respectively (figure \ref{fig:RDF_mod}(a)). The density peak is also observed at larger distances for bigger simulated fish. The repulsive force is practically the same for all three cases at any given distance (figure \ref{fig:RDF_mod}(c)) as the radii of the repulsion zone are constant for all fish sizes. The attractive portion of the $F(r)$-curves has a complex shape as in the experiment indicating strongest attraction at $r \approx 28$, $r\approx90$ and $r\approx200$ mm for small, medium and large simulated fish, respectively.
For all five group sizes of the medium sized simulated fish ($k=1.73$) the maximum of the local density is well above the system's average (figure \ref{fig:RDF_mod}(d)) and is largest for groups of 10 fish ($g(r)=52$). All the curves cross the $g(r)=1$ line in the same order as in the experiment: The average half-radial width of groups of 10 and 60 simulated fish are $\sim$ 100 mm and $\sim$ 210 mm, respectively. As in the experiment, the repulsive force (figure \ref{fig:RDF_mod}(f)) decays with a distance and takes very similar values for all five cases. The attraction in groups of 10 and 20 individuals is maximal at $r=90$ and $r=115$, respectively, then decreases steeply with a distance and reaches zero at $r\approx190$. For the other three curves (N=30, 40 and 60), the attractive force has a maximum at $r\approx100$. At all intermediate distances the dissipation is stronger in small groups of simulated fish which is in agreement with experiments.

\section{Discussion}
Although many recent studies have focused on understanding interactions between fish in schools \cite{herbert-read.j:2011,katz_inferring_2011,tunstrom.k:2013}, a detailed description of large systems consisting of tens or hundreds of individuals thus far remained a challenge. In analysis of such systems, the quality of fish tracking becomes a limiting factor, as many of the techniques traditionally used to study fish behaviour require high consistency of individual fish identities over time in order to reconstruct velocity or acceleration profiles of the individuals. When the long-time individual identities are not available, an approach that relies only on individuals' positional data, such as Boltzmann inversion, can be useful for assessing interactions in large groups. The latter approach is also faster as it works with spatial data only, and the pair distribution function used in its first step allows also to extract useful information about aggregations and clustering in the system. Such approach can also potentially simplify fish tracking and post-processing of experimental data.

This approach, however, also has its limitations. The direct Boltzmann inversion has become a popular method to derive interaction potentials across different fields first of all due to its simplicity and general applicability. Even though this method has a straightforward nature, the potentials derived with the Boltzmann inversion are generally non-unique and can be state-dependent. They can also be influenced by long-range correlations or by the anisotropic nature of the interactions that is inherent to many living systems. Moreover, these potentials are effective as they may include multi-particle correlations (e.g. simultaneous response to positioning or movements of several conspecifics) and correlated contributions from the surroundings, such as geometrical confinement effects. Such potentials thus do not share the typical properties attached to equilibrium potentials and this limits their use mostly to qualitative description of the interactions. This also suggests that to draw a conclusive quantitative picture on how individuals interact in a group, a method like Boltzmann inversion needs to be accompanied by rigorous analyses of other motion statistics and/or by extensive computer simulations of the system of interest. 

The challenges of using self-propelled particle models for modeling collective animal behaviour are also known \cite{sumpter.djt:2010,herbert-read.j:2015}. The tractable models are usually oversimplified and too general and thus lack flexibility required to reflect the features of a particular phenomena. In this work, we tried to overcome many of these limitations by parameterising our simple SPP model with experimental data. Some of the simplifications, on the other hand, were introduced in the model deliberately, such as the absence of an explicit attraction rule for the interaction between the agents (see also Model section). We found the inclusion of the attraction rule unnecessary since a combination of fish-fish alignment and fish-wall interactions proved to be sufficient to reproduce the dynamics observed in experiments both statistically and visually \cite{herbert-read.j:2015}. We should note, however, that the effective cohesion detected in the model system by the direct Boltzmann inversion method is hence attributed to a combination of the inter-individual and agent-wall interactions. Even more detailed description of the system and possibly a better fit to experimental data could be achieved if mass and inertial forces of the individuals are taken into account. These features can be relatively easily implemented in a model which describes the motion of individuals by including friction and stochastic forces, e.g. Active Brownian Particle model with aligning interactions \cite{lobaskin.v:2013,grossmann12}.

\section{Conclusion}
Using a combination of Boltzmann inversion and traditional statistical methods, we have inferred how fish ({\it Pseudomugil signifer}) interact in schools. Previous studies have applied a force-matching approach to infer the interactions of schooling fish from their movements \cite{katz_inferring_2011}. Our method can infer these interactions directly from the static spatial distribution of individuals in groups. Whilst repulsion forces had the same strength for different sized fish, attraction strength increased in larger fish, consistent with how a fish's movement develops with age \cite{masuda.r:1998}. The interactions between fish also changed as a function of group size, as suggested by other studies \cite{gautrais.j:2012}. Our model, refined on the basis of these observations, could capture the dynamics of schooling fish. These findings are also in line with the results of our previous study \cite{herbert-read.j:2015}, where we utilised an observational test to cross-validate the model used in the present study. We expect that our findings could also generalise to other species that exhibit schooling behaviour \cite{Katz198120,OBRIEN19891}. Application of the approaches used in statistical physics, coupled with informed models of collective motion, now allows us to shed more light on the intricacies of how individuals in groups interact.  

\subsubsection*{Data accessibility}{The datasets supporting this article are available from http://dx.doi.org/10.5061/dryad.8g0d0}

\subsubsection*{Authors' contributions}{M.R., J.E.H.-R, D.J.T.S. and A.J.W.W. conceived/designed the study and wrote the paper. J.E.H.-R. and A.J.W.W. designed the experiments. J.E.H.-R. performed the experiments and tracking. M.R. analysed the data, developed the computational model, performed the simulations and prepared figures. All authors gave final approval for publication.}

\subsubsection*{Competing interests}{We have no competing interests.}

\subsubsection*{Ethics statement.}{Investigations were performed under ethical permission from the University of Sydney's Ethics Committee (ref. number: L04/6- 2009/3/5083).}

\subsubsection*{Acknowledgment}{M.R. would like to thank Vladimir Lobaskin for valuable discussions. The authors thank Alexander Szorkovszky for his constructive comments on the manuscript.}

\subsubsection*{Funding}{This work is supported by a Knut and Alice Wallenberg foundation grant No. 2013.0072 to D.J.T.S.}

\vfil

\widetext
\clearpage
\begin{center}
\textbf{\large Supplemental Material for: Body size affects the strength of social interactions and spatial organisation of a schooling fish ({\it Pseudomugil signifer})}
\end{center}
\begin{center}
\text{Maksym Romenskyy,\textsuperscript{1} James E. Herbert-Read,\textsuperscript{1,2} Ashley J. W. Ward,\textsuperscript{3} David J. T. Sumpter\textsuperscript{1}}
\end{center}
\begin{center}
\textit{\textsuperscript{1}Department of Mathematics, Uppsala University, Box 480, Uppsala 75106, Sweden}
\textit{\textsuperscript{2}Department of Zoology, Stockholm University, Stockholm 10691, Sweden}
\textit{\textsuperscript{3}School of Biological Sciences, University of Sydney, Sydney, New South Wales, Australia}
\end{center}
\begin{center}
\text{(Dated: \today)}
\end{center}
\setcounter{equation}{0}
\setcounter{figure}{0}
\setcounter{table}{0}
\setcounter{page}{1}
\makeatletter
\renewcommand{\theequation}{S\arabic{equation}}
\renewcommand{\thefigure}{S\arabic{figure}}
\renewcommand{\bibnumfmt}[1]{[S#1]}
\renewcommand{\citenumfont}[1]{S#1}

\section{Experimental details}

\subsection{Materials}

\begin{figure}[h!]
\centering
\subfigure{
\includegraphics[width=7.75cm,clip]{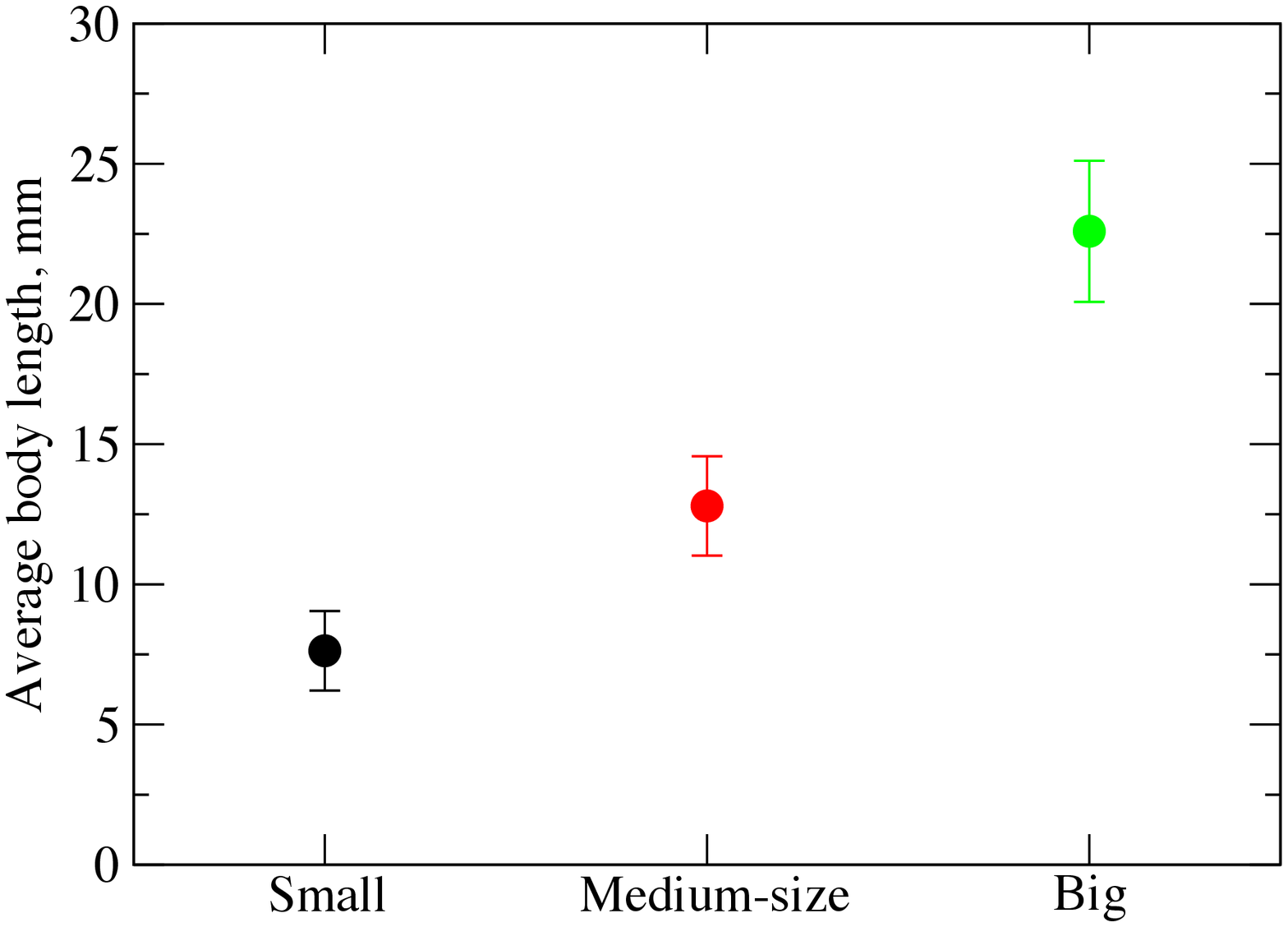}
}
\subfigure{
\includegraphics[width=8.3cm,clip]{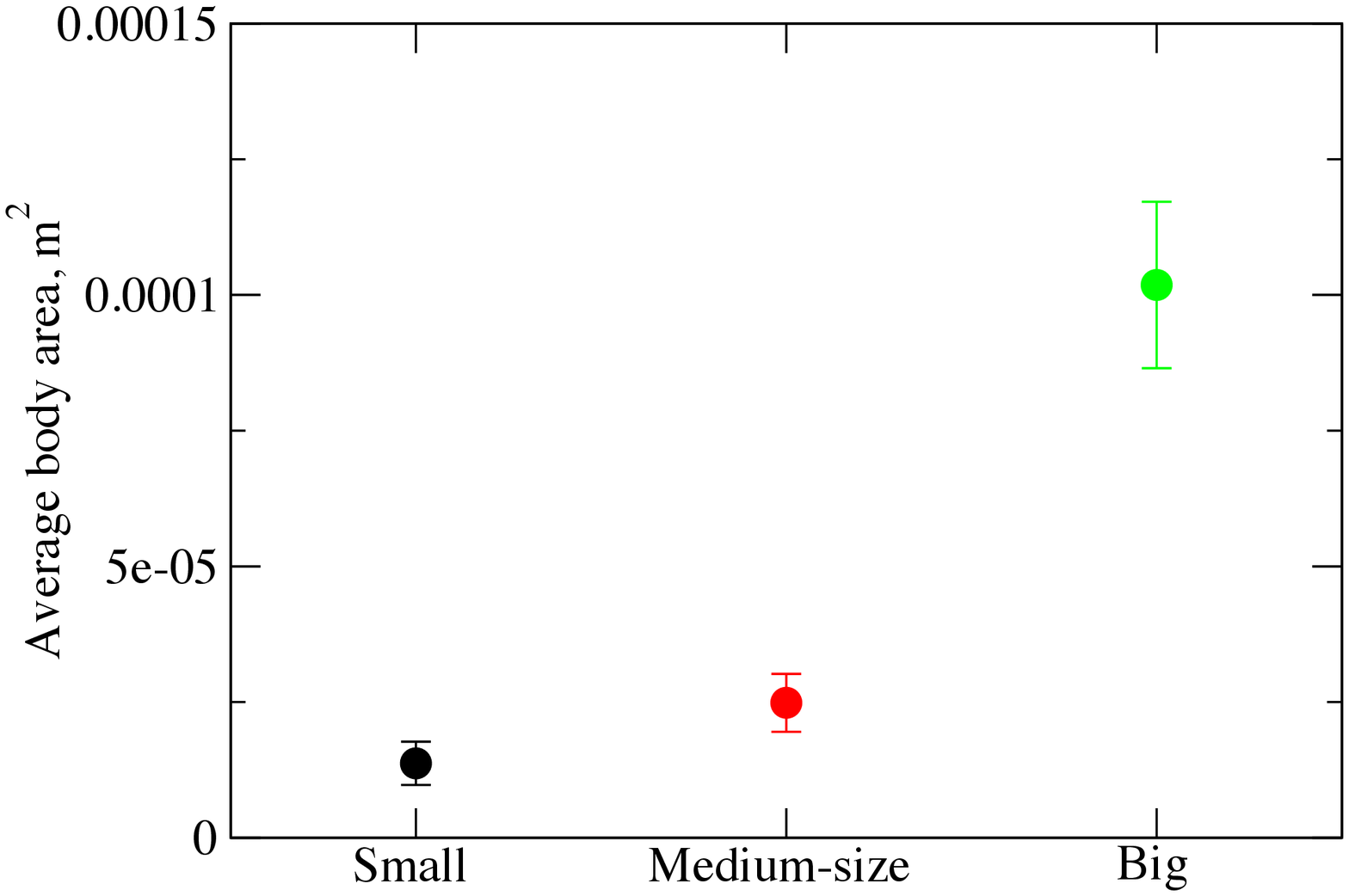}
}
\caption{Average fish body length $\pm1$ s.d. for each size class (left). Average fish body area $\pm1$ s.d. for each size class (right).}
\label{fig:bl}
\end{figure}

Pacific blue-eye fish ({\it Pseudomugil signifer}) were caught in hand nets from Narrabeen Lagoon, New South Wales, Australia ($33^\circ43^\prime03$ S, $151^\circ16^\prime17$ E). Fish were kept in filtered freshwater in 150 l glass tanks at $22-25^\circ$ and fed crushed flake food {\it ad libitum}. All fish were housed for at least three weeks prior to experimentation. The experimental arena was circular with a diameter of 760 mm. It was filled to a depth of 70 mm with aged and conditioned tap water. The arena was lit by fluorescent lamps and was visually isolated. For each trial, we randomly selected $N$ fish ($N=10, 20, 30, 40, 50$ or $60$ for small and medium fish and $N=10, 20, 30, 40$ for large fish) of similar size (see Fig. \ref{fig:bl}) from the housing tanks and placed them in the experimental arena. Fish were left to acclimate to the arena for at least five minutes, after which they were filmed for 15-20 minutes at 15 frames per second. We used a Logitech Pro 9000 camera placed orthogonally to the arena above its geometrical centre at a distance $>1$ m minimising the radial distortion. The number of trials for each group size ranged between 3-10 (see Table \ref{nvideos}) due to limitations in the number of fish we could obtain for large or small body sizes. Because of the large numbers of fish we used for the experiment, we reused fish between trials. Fish were never used more than once per day and fish were used a maximum of 5 times.  

\begin{table}
\centering
\begin{tabular}{|c|c|c|c|}
\hline
Group size & \multicolumn{3}{c|}{Average body length} \\ \cline{2-4} 
                            & 7.5 mm    & 13 mm    & 23 mm   \\ \hline
10                          & 4         & 10        & 3       \\ \hline
20                          & 4         & 8        & 3       \\ \hline
30                          & 4         & 8        & 3      \\ \hline
40                          & 3         & 9        & 3       \\ \hline
50                          & 3         & 6        & --       \\ \hline
60                          & 3         & 6        & --       \\ \hline
\end{tabular}
\caption{Number of videos recorded for each group size and body size. Each film is 15-20 min long.%
\label{nvideos}}
\end{table}

\subsection{Data collection and acquisition}

Films were recorded in .mov format using original camera manufacturer software and subsequently converted to .avi using DirectShowSource and VirtualDub (v 1.9.2). The tracking was performed using DIDSON tracking program \cite{handegard.n:2008}. The raw data consisted of $x$ and $y$ coordinates, fish identity and a time stamp. The accuracy of the tracking process was checked by projecting the raw tracking data onto experimental videos.

Figure \ref{fig:RD} shows radial distribution of fish in the arena calculated as $g(R)=\rho(R)/\langle \rho \rangle$, where $\rho(R)$ is the density of fish in the circular shell of mean radius $R$ and $\langle \rho \rangle$ is the average fish density in the system. Small fish $l=7.5$ mm are distributed more regularly throughout the tank as compared to large individuals $l=23$ mm tending to move close to the arena wall.

\section{Motion statistics analysis}

\subsection{Calculation of the shell area for the pair distribution function in confinement}

\begin{figure}
\centering
\includegraphics[width=8.0cm,clip]{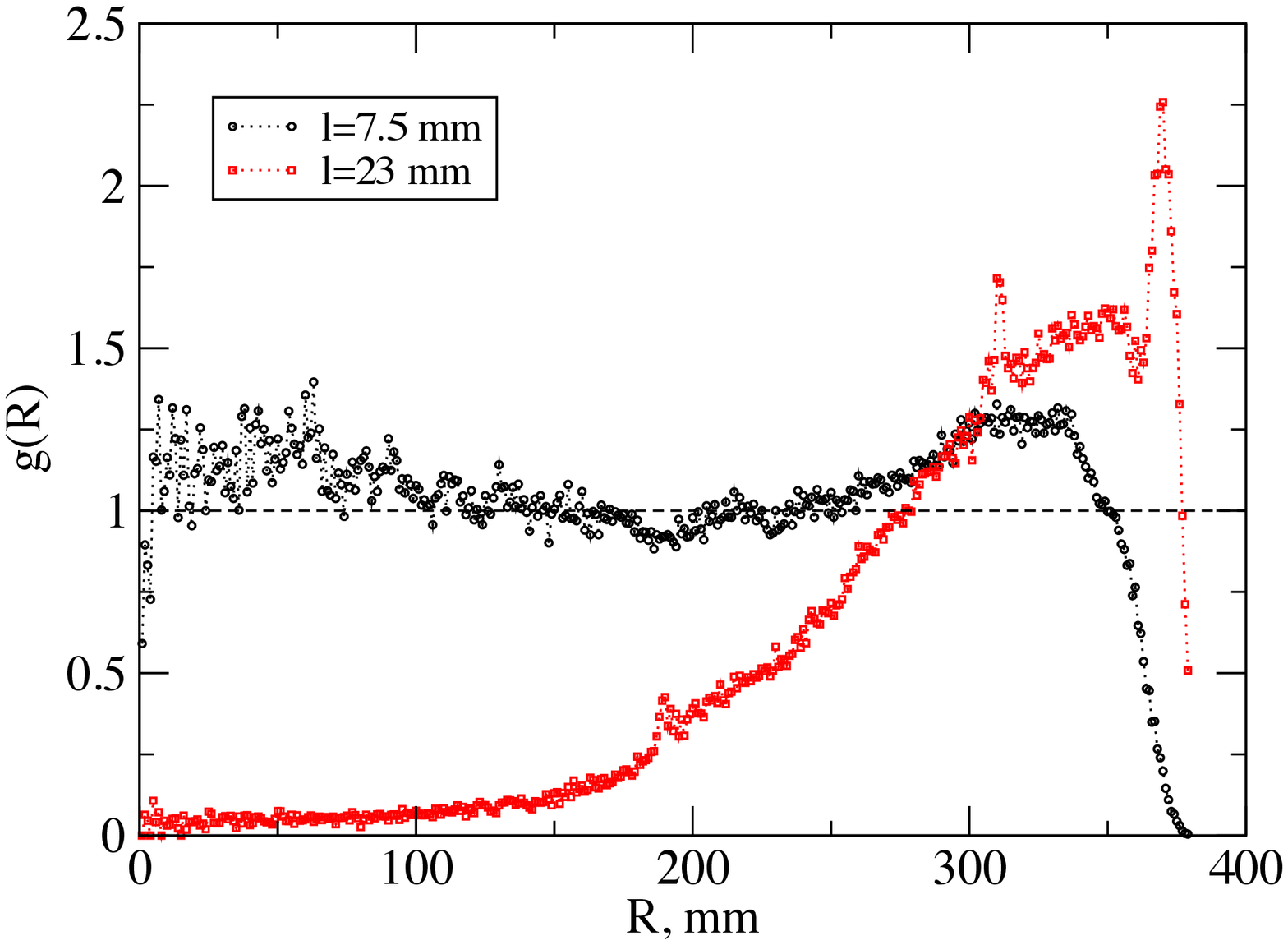}
\caption{Distribution of fish across the arena in radial direction ($N=40$). 0 mm corresponds to the center of the arena; the wall is located at 380 mm from the center. The dotted black line at $g(R)=1$ corresponds to completely homogeneous distribution of fish throughout the arena.}
\label{fig:RD}
\end{figure}

\begin{figure}
\centering
\includegraphics[width=7.0cm,clip]{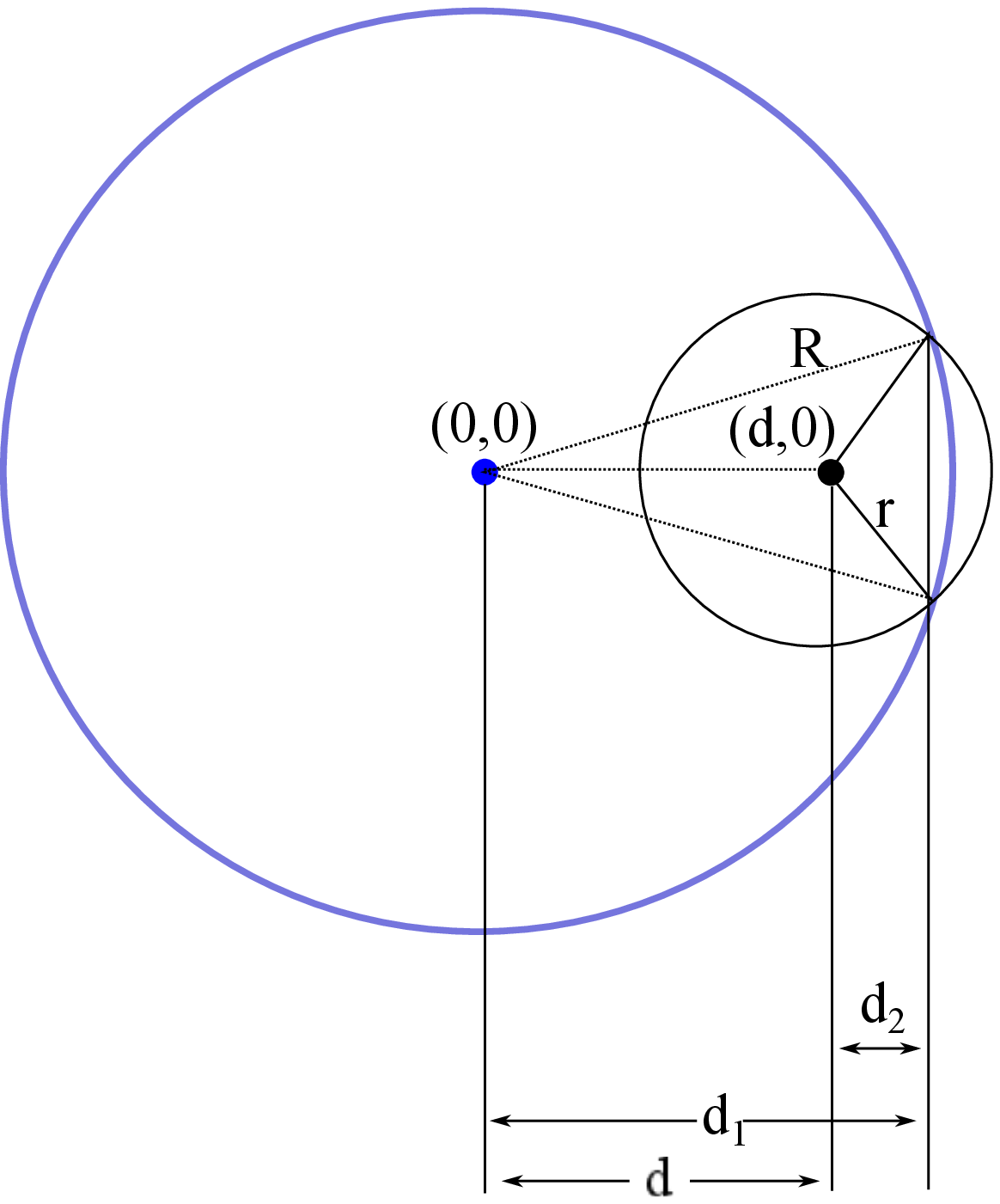}
\caption{Illustration of the intersection of the pair distribution function circle $r$ with confining circle $R$. Only one circle for the pair distribution function is drawn for simplicity.}
\label{fig:RDF_circles}
\end{figure}

When calculating the pair distribution function for confined systems particular care needs to be taken when the particles are located close to the wall, at a distance smaller than the radius of the largest shell. In experiments with fish this scenario is common (see Fig. \ref{fig:RD} for illustration). In such situations only area of the shell lying inside the boundaries of the confining geometry should be considered. For all cases the shell area can be calculated as a difference between the areas of two neighbouring circles. The two latter areas for convenience can be computed separately as intersections excluding the areas outside the constraint. If confinement is circular (see Fig. \ref{fig:RDF_circles}), as is in our case, to find the intersection area we can use a formula for the circular segment of triangular height $d'$ (excluding the height of the arced portion) and radius $R'$ \cite{weinstein.ew}:
\begin{equation}
S(R',d')={R'}^2\cos^{-1}(d'/R')-d'\sqrt{R'^2-d'^2},
\tag{S1}\label{A_r_d}
\end{equation}

In the simplified case that we consider here (with only one circle for the pair distribution function) we have two such segments, one for the circular confinement (experimental arena) with a radius $R$ and hight  $d_1$, and another one for the shell boundary of the pair distribution function, having radius $r$ and height $d_2$. The heights are calculated as

\begin{equation}
d_1=\frac{d^2-r^2+R^2}{2d},
\tag{S2}\label{d_1}
\end{equation}
and
\begin{equation}
d_2=\frac{d^2+r^2-R^2}{2d}.
\tag{S3}\label{d_2}
\end{equation}

To calculate the total area of the intersection Eq. \eqref{A_r_d} needs to be solved two times, once for each segment. Thus, combining Eqs. (\ref{A_r_d}-\ref{d_2}) we get
\begin{widetext}
\begin{equation}
\begin{split}
S(r')&=S(R,d_1)+S(r,d_2)=\\
&=r^2\cos^{-1}\left(\frac{d^2+r^2-R^2}{2dr}\right)+R^2\cos^{-1}\left(\frac{d^2+R^2-r^2}{2dR}\right)-\frac{1}{2}\sqrt{(-d+r+R)(d+r-R)(d-r+R)(d+r+R)}.
\end{split}
\tag{S4}\label{area_c}
\end{equation}
\end{widetext}

\begin{figure}
\centering
\includegraphics[width=8.0cm,clip]{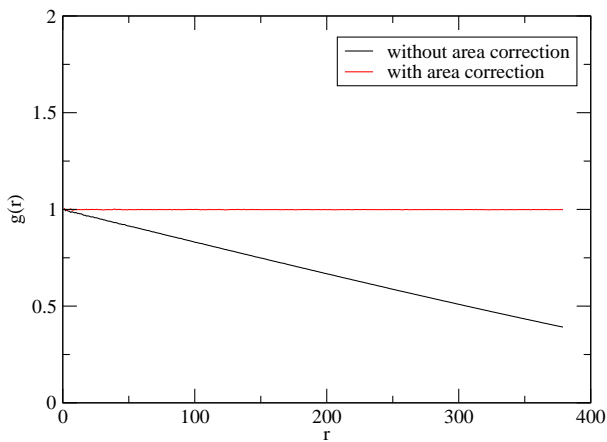}
\caption{Pair distribution function $g(r)$ for a homogeneous test system constituting of 1000 particles (average over $5\times10^6$ positional configurations), $R=380$.}
\label{fig:RDF_correct}
\end{figure}

Figure \ref{fig:RDF_correct} shows a plot of the pair distribution of particles for highly homogeneous system in circular confinement (distribution of particles is uniform). The black curve displays a clear linear decay of $g(r)$ with increasing inter-particle separation distance $r$. This represents the case where the area outside the constraint has also been included in calculations. The red curve represents the case when all the shell areas have been calculated with the method described above. This shows highly regular distribution $g(r)$ of particles for all separation distances $r$ ($1:1$ relation between the local shell density and average density in the system).

\subsection{Calculation of the surface area of a fish group and the body area of an individual fish}

\begin{figure}
\centering
\subfigure{
\includegraphics[width=4.1cm,clip]{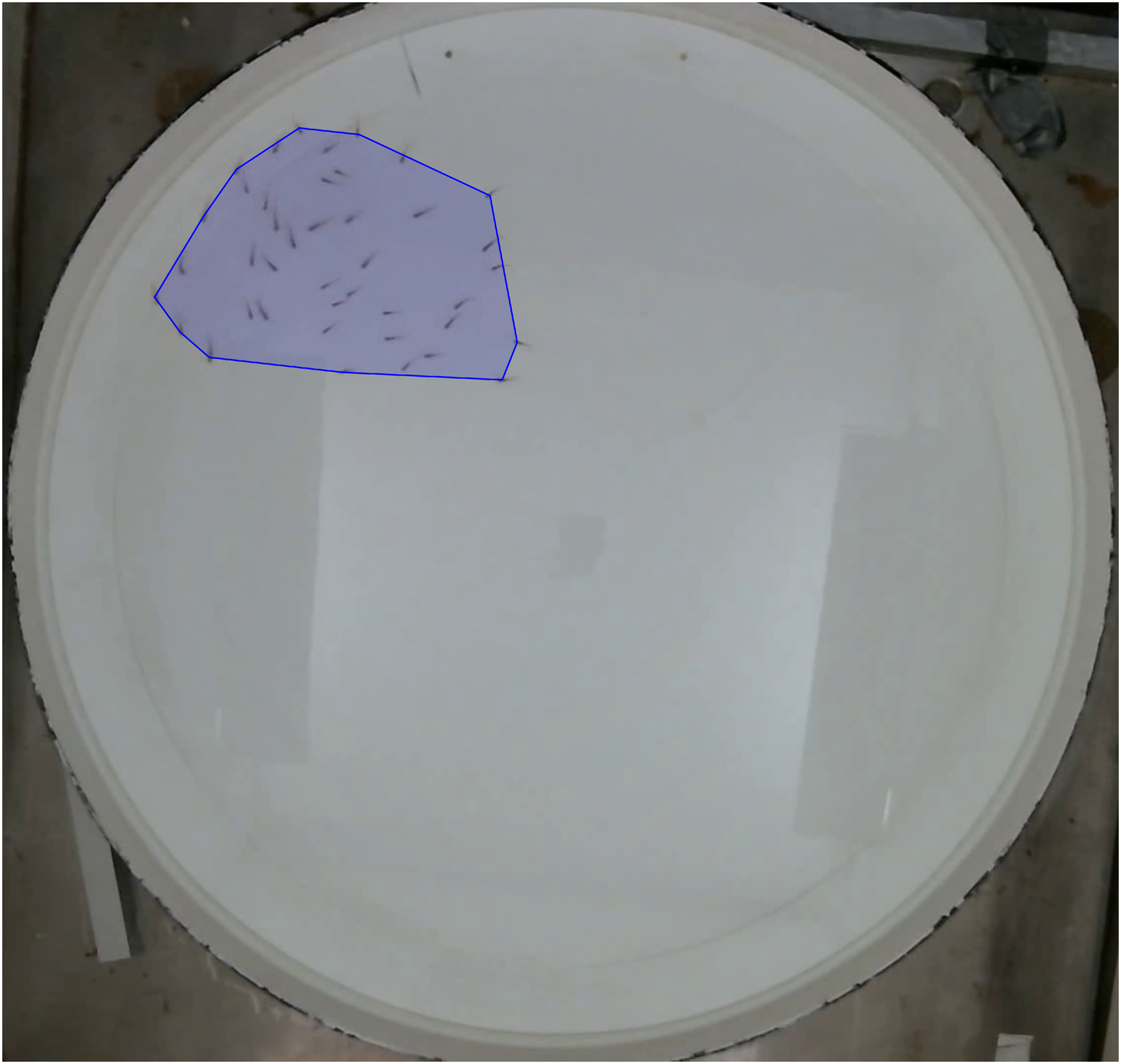}
}
\subfigure{
\includegraphics[width=3.9cm,clip]{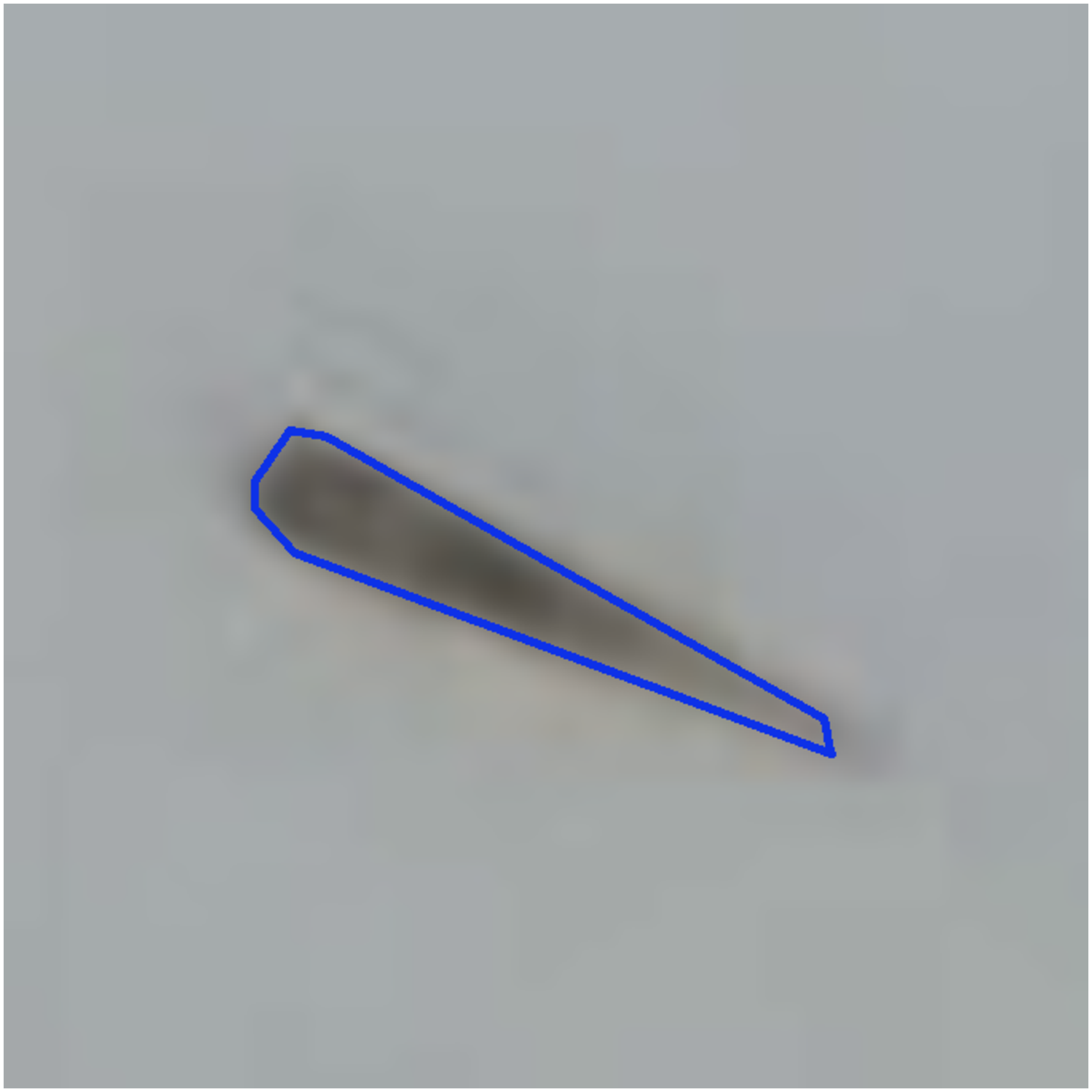}
}
\caption{Complex shapes obtained from the experimental data. Left: Typical polygon returned by a convex hull algorithm. Right: Fish body shape estimated as a convex shape. The displayed frame is taken from a video for 40 medium-size fish. The frame on the left has been scaled for clarity purposes.}
\label{fig:ch_area}
\end{figure}

The surface area of a group was calculated for every frame of a video. First, we computed a convex hull $C$ of a set of $N$ points representing geometrical centres of fish bodies \cite{Barber:1996}. The convex hull is defined by

\begin{equation}
C\equiv{\sum^{N}_{i=1}\lambda_ip_i:\lambda\geq0} \mbox{ for all } i \mbox{ and } {\sum^{N}_{i=1}\lambda_i=1},
\tag{S5}\label{convex_hull}
\end{equation}

where $i$ is the point (fish) index of a point $p_i$ and $\lambda$ is the non-negative weight coefficient. The resulting convex hull gives the identities of the vertices of a polygon as an output. The area $A$ of this polygon can be computed as \cite{bourke.p}

\begin{equation}
A=\frac{1}{2}\sum^{N}_{i=1}{(x_iy_{i+1}-x_{i+1}y_i)},
\tag{S6}\label{area_group}
\end{equation}

where $x_i$ and $y_i$ represent coordinates of the vertices of a polygon. The last vertex $(x_{N+1},y_{N+1})$ is assumed to be the same as the first one, so $(x_{N+1},y_{N+1}) =(x_1,y_1)$ and the polygon is closed.

The body area of an individual fish was computed for every identified individual in every frame of a movie. All edge pixels forming a shape of an object (fish) were identified based on the preset weighted intensity threshold. The resulting shape was a polygon with N vertices corresponding to a number of edge pixels with intensity values above the threshold. The area of this complex shape was computed with the same method as for the surface area of a group.

\subsection{Additional motion statistics}

\begin{figure}
\centering
\includegraphics[width=9cm,clip]{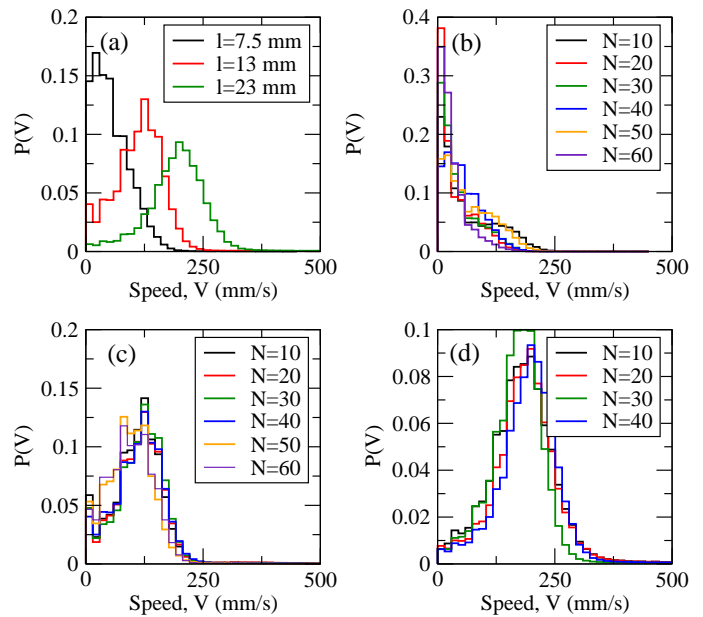}
\caption{Experimental speed distributions for: (a) three sizes of fish in groups of 40 individuals, (b) small (l=7.5 mm), (c) medium (l=13 mm) and (d) large (l=23 mm) fish at variable group size.}
\label{fig:vel_all}
\end{figure}

Figure \ref{fig:vel_all}(a) shows speed distributions for fish of three average body lengths at fixed group size (N=40). Small fish (l=7.5 mm) most of the time have speeds below 100 mm/s and very often within a range of 0-20 mm/s. Very rarely small fish have speed above 200 mm/s. For medium-size individuals the distribution is much wider and has a peak at $V=125$ mm/s. It spans up to $V\approx 250$ mm/s and has another minimum at 0-20 mm/s. For large fish the number of events when the individuals are stationary or barely move decreases further and the peak is observed at $V\approx 200$. Figures \ref{fig:vel_all}(b)-(d) show speed distributions for small, medium and large fish and various group sizes. All histograms for the same body size practically overlap. Therefore, while speed regime of fish depends strongly on body size it is not effected by the number of individuals in a group.

\begin{figure}
\centering
\includegraphics[width=9cm,clip]{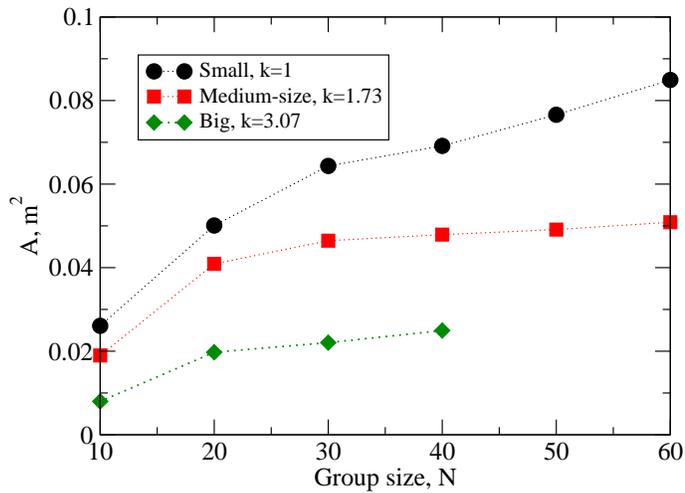}
\caption{Average area of a simulated group of fish for three size classes: small ($k=1$), medium-size ($k=1.73$), and big ($3.07$) fish. Simulation parameters as stated in the main text, Methods section.}
\label{fig:area_sim}
\end{figure}

Figure \ref{fig:area_sim} shows the average group area occupied by simulated fish as a function of number of individuals in a group. For all three sizes of simulated fish, the group area increases with group size, in agreement with experimental results (Fig. 3(a), main text). For all group sizes, big fish form densest groups. Conversely, groups of small fish occupy much larger area as compared to the other two size classes (medium-size and big fish).


\begin{thebibliography}{10}

\bibitem{Romensky20150015}
Maksym Romensky, Dimitri Scholz, and Vladimir Lobaskin.
\newblock Hysteretic dynamics of active particles in a periodic orienting
  field.
\newblock {\em J. Royal Soc. Interface}, 12(108), 2015.

\bibitem{vicsek.t:2012}
T.~Vicsek and A.~Zafeiris.
\newblock Collective motion.
\newblock {\em Phys. Rep.}, 517:71, 2012.

\bibitem{herbert-read.j:2011}
J.E. Herbert-Read, A.~Perna, R.P. Mann, T.M. Schaerf, D.J.T. Sumpter, and
  A.J.W. Ward.
\newblock Inferring the rules of interaction of shoaling fish.
\newblock {\em Proc. Natl. Acad. Sci. USA}, 108(46):18726--18731, 2011.

\bibitem{herbert2015initiation}
James~E Herbert-Read, Jerome Buhl, Feng Hu, Ashley~JW Ward, and David~JT
  Sumpter.
\newblock Initiation and spread of escape waves within animal groups.
\newblock {\em Royal Soc. Open Science}, 2(4):140355, 2015.

\bibitem{bialek.w:2011}
W.~Bialek, A.~Cavagna, I.~Giardina, T.~Morad, E.~Silvestri, M.~Viale, and A.M.
  Walczak.
\newblock Statistical mechanics for natural flocks of birds.
\newblock {\em Proc. Natl. Acad. Sci. USA}, 109(13):4786--4791, 2011.

\bibitem{toner.j:2005}
J.~Toner, Y.~Tu, and S.~Ramaswami.
\newblock Hydrodynamics and phases of flocks.
\newblock {\em Ann. Phys.}, 318:170--244, 2005.

\bibitem{buhl.j:2006}
J.~Buhl, D.~J.~T. Sumpter, I.~D. Couzin, J.~J. Hale, E.~Despland, E.~R. Miller,
  and S.~J. Simpson.
\newblock From disorder to order in marching locusts.
\newblock {\em Science}, 312(5778):1402--1406, 2006.

\bibitem{calovi.ds:2014}
Daniel~S Calovi, Ugo Lopez, Sandrine Ngo, Cl\'{e}ment Sire, Hugues Chat\'{e},
  and Guy Theraulaz.
\newblock Swarming, schooling, milling: phase diagram of a data-driven fish
  school model.
\newblock {\em New. J. Phys.}, 16(1):015026, 2014.

\bibitem{tunstrom.k:2013}
Kolbj\o{}rn Tunstr\o{}m, Yael Katz, Christos~C. Ioannou, Cristi\'an Huepe,
  Matthew~J. Lutz, and Iain~D. Couzin.
\newblock Collective states, multistability and transitional behavior in
  schooling fish.
\newblock {\em PLoS Comput. Biol.}, 9(2):e1002915, 02 2013.

\bibitem{attanasi.a:2014}
A~Attanasi, A~Cavagna, L~Del~Castello, Irene Giardina, Stefania Melillo,
  Leonardo Parisi, Oliver Pohl, Bruno Rossaro, Edward Shen, Edmondo Silvestri,
  and Massimiliano Viale.
\newblock Collective behaviour without collective order in wild swarms of
  midges.
\newblock {\em PLoS Comput. Biol.}, 10(7):e1003697, 2014.

\bibitem{cavagna.a:2010}
Andrea Cavagna, Alessio Cimarelli, Irene Giardina, Giorgio Parisi, Raffaele
  Santagati, Fabio Stefanini, and Massimiliano Viale.
\newblock Scale-free correlations in starling flocks.
\newblock {\em Proc. Natl. Acad. Sci. USA}, 107(26):11865--11870, 2010.

\bibitem{bialek.w:2014}
William Bialek, Andrea Cavagna, Irene Giardina, Thierry Mora, Oliver Pohl,
  Edmondo Silvestri, Massimiliano Viale, and Aleksandra~M. Walczak.
\newblock Social interactions dominate speed control in poising natural flocks
  near criticality.
\newblock {\em Proc. Natl. Acad. Sci. USA}, 111(20):7212--7217, 2014.

\bibitem{cavagna.a:2013}
Andrea Cavagna, S.~M. Duarte~Queir\'{o}s, Irene Giardina, F.~Stefanini, and
  M.~Viale.
\newblock Diffusion of individual birds in starling flocks.
\newblock {\em Proc. R. Soc. B}, 280(1756):20122484, 2013.

\bibitem{cavagna.a:2014}
Andrea Cavagna and Irene Giardina.
\newblock Bird flocks as condensed matter.
\newblock {\em Ann. Rev. Cond. Matt. Phys.}, 5(1):183--207, 2014.

\bibitem{narayan.v:2007}
V.~Narayan, S.~Ramaswamy, and N.~Menon.
\newblock Long-lived giant number fluctuations in a swarming granular nematic.
\newblock {\em Science}, 317:105--108, 2007.

\bibitem{cavagna.a1:2014}
Andrea Cavagna, Irene Giardina, Francesco Ginelli, Thierry Mora, Duccio
  Piovani, Raffaele Tavarone, and Aleksandra~M. Walczak.
\newblock Dynamical maximum entropy approach to flocking.
\newblock {\em Phys. Rev. E}, 89:042707, Apr 2014.

\bibitem{meckelke.m:2013}
Martin Mechelke and Michael Habeck.
\newblock Estimation of interaction potentials through the configurational
  temperature formalism.
\newblock {\em J. Chem. Theory Comp.}, 9(12):5685--5692, 2013.

\bibitem{moore.t:2014}
Timothy~C. Moore, Christopher~R. Iacovella, and Clare McCabe.
\newblock Derivation of coarse-grained potentials via multistate iterative
  boltzmann inversion.
\newblock {\em J. Chem. Phys.}, 140(22):--, 2014.

\bibitem{mcgreevy.r:1988}
R.~L. McGreevy and L.~Pusztai.
\newblock Reverse monte carlo simulation: A new technique for the determination
  of disordered structures.
\newblock {\em Mol. Simulat.}, 1(6):359--367, 1988.

\bibitem{lyubartsev.a:1995}
Alexander~P. Lyubartsev and Aatto Laaksonen.
\newblock Calculation of effective interaction potentials from radial
  distribution functions: A reverse monte carlo approach.
\newblock {\em Phys. Rev. E}, 52:3730--3737, Oct 1995.

\bibitem{soper.a:1996}
A.K. Soper.
\newblock Empirical potential monte carlo simulation of fluid structure.
\newblock {\em Chem. Phys.}, 202(2-3):295 -- 306, 1996.

\bibitem{tschop.w:1998}
W.~Tsch\"{o}p, K.~Kremer, J.~Batoulis, T.~B\"{u}rger, and O.~Hahn.
\newblock Simulation of polymer melts. i. coarse-graining procedure for
  polycarbonates.
\newblock {\em Acta Polymerica}, 49(2-3):61--74, 1998.

\bibitem{couzin.id:2002}
I.~D. Couzin, J.~Krause, R.~James, G.~D. Ruxton, and N.~R. Franks.
\newblock Collective memory and spatial sorting in animal groups.
\newblock {\em J. Theor. Biol.}, 218:1--11, 2002.

\bibitem{Hinz13022017}
Robert~C. Hinz and Gonzalo~G. de~Polavieja.
\newblock Ontogeny of collective behavior reveals a simple attraction rule.
\newblock {\em Proceedings of the National Academy of Sciences}, 2017.

\bibitem{masuda.r:1998}
R.~Masuda and K.~Tsukamoto.
\newblock The ontogeny of schooling behaviour in the striped jack.
\newblock {\em J. Fish Biol.}, 52(3):483--493, 1998.

\bibitem{pusey.b:2004}
Brad Pusey, Mark Kennard, and Angela Arthington.
\newblock {\em Freshwater Fishes of North-Eastern Australia}.
\newblock Collingwood, Victoria: Csiro Publishing, 2004.

\bibitem{handegard.n:2008}
Nils~Olav Handegard and Kresimir Williams.
\newblock Automated tracking of fish in trawls using the didson (dual frequency
  identification sonar).
\newblock {\em ICES J. Marine Sci.: Journal du Conseil}, 65(4):636--644, 2008.

\bibitem{romenskyy.m:2013}
M.~Romenskyy and V.~Lobaskin.
\newblock Statistical properties of swarms of self-propelled particles across
  the order-disorder transition.
\newblock {\em Eur. Phys. J. B}, 86:91, 2013.

\bibitem{vicsek.t:1995}
T.~Vicsek, A.~Czir\'{o}k, E.~Ben-Jacob, I.~Cohen, and O.~Shochet.
\newblock Novel type of phase transition in a system of self-driven particles.
\newblock {\em Phys. Rev. Lett.}, 75:1226--1229, 1995.

\bibitem{li.w:2007}
Wei Li and Xiaofan Wang.
\newblock Adaptive velocity strategy for swarm aggregation.
\newblock {\em Phys. Rev. E}, 75:021917, Feb 2007.

\bibitem{zhang.j:2009}
Jue Zhang, Yang Zhao, Baomei Tian, Liqian Peng, Hai-Tao Zhang, Bing-Hong Wang,
  and Tao Zhou.
\newblock Accelerating consensus of self-driven swarm via adaptive speed.
\newblock {\em Physica A: Statistical Mechanics and its Applications},
  388(7):1237 -- 1242, 2009.

\bibitem{mishra.s:2012}
Shradha Mishra, Kolbj\o{}rn Tunstr\o{}m, Iain~D. Couzin, and Cristi\'an Huepe.
\newblock Collective dynamics of self-propelled particles with variable speed.
\newblock {\em Phys. Rev. E}, 86:011901, Jul 2012.

\bibitem{lu.s:2013}
Shengtao Lu, Wuguo Bi, Fang Liu, Xiangyang Wu, Bengang Xing, and Edwin K.~L.
  Yeow.
\newblock Loss of collective motion in swarming bacteria undergoing stress.
\newblock {\em Phys. Rev. Lett.}, 111:208101, Nov 2013.

\bibitem{gautrais.j:2009}
Jacques Gautrais, Christian Jost, Marc Soria, Alexandre Campo, S\'{e}bastien
  Motsch, Richard Fournier, St\'{e}phane Blanco, and Guy Theraulaz.
\newblock Analyzing fish movement as a persistent turning walker.
\newblock {\em J. Math. Biol.}, 58(3):429--445, 2009.

\bibitem{gautrais.j:2012}
Jacques Gautrais, Francesco Ginelli, Richard Fournier, St\'{e}phane Blanco,
  Marc Soria, Hugues Chat\'{e}, and Guy Theraulaz.
\newblock Deciphering interactions in moving animal groups.
\newblock {\em PLoS Comput. Biol.}, 8(9):e1002678, 09 2012.

\bibitem{domenici.p:1997}
P.~Domenici and R.~W. Blake.
\newblock The kinematics and performance of fish fast-start swimming.
\newblock {\em J. Experimental Biol.}, 200:1165--1178, 1997.

\bibitem{ballerini.m:2008}
Michele Ballerini, Nicola Cabibbo, Raphael Candelier, Andrea Cavagna, Evaristo
  Cisbani, Irene Giardina, Alberto Orlandi, Giorgio Parisi, Andrea Procaccini,
  Massimiliano Viale, and Vladimir Zdravkovic.
\newblock Empirical investigation of starling flocks: a benchmark study in
  collective animal behaviour.
\newblock {\em Anim. Behav.}, 76(1):201 -- 215, 2008.

\bibitem{cavagna.a:2008}
Andrea Cavagna, Alessio Cimarelli, Irene Giardina, Alberto Orlandi, Giorgio
  Parisi, Andrea Procaccini, Raffaele Santagati, and Fabio Stefanini.
\newblock New statistical tools for analyzing the structure of animal groups.
\newblock {\em Math. Biosci.}, 214(12):32 -- 37, 2008.

\bibitem{peruani.f:2012}
F.~Peruani, J.~Starru\ss, V.~Jakovlevic, L.~S{\o}gaard-Andersen, A.~Deutsch,
  and M.~B\"{a}r.
\newblock Collective motion and nonequilibrium cluster formation in colonies of
  gliding bacteria.
\newblock {\em Phys. Rev. Lett.}, 108:098102, 2012.

\bibitem{dinsmore.ad:1997}
A.D. Dinsmore, P.B. Warren, W.~C.~K. Poon, and A.~G. Yodh.
\newblock Fluid-solid transitions on walls in binary hard-sphere mixtures.
\newblock {\em Europhys. Lett.}, 40(3):337--342, 1997.

\bibitem{pearce}
D.~J.~G. Pearce and M.~S. Turner.
\newblock Emergent behavioural phenotypes of swarming models revealed by
  mimicking a frustrated anti-ferromagnet.
\newblock {\em Journal of The Royal Society Interface}, 12(111), 2015.

\bibitem{katz_inferring_2011}
Yael Katz, Kolbj\"orn Tunstr\"om, Christos~C. Ioannou, Cristián Huepe, and
  Iain~D. Couzin.
\newblock Inferring the structure and dynamics of interactions in schooling
  fish.
\newblock {\em P Natl. Acad. Sci.}, 108(46):18720--18725, November 2011.

\bibitem{sumpter.djt:2010}
D.~J.~T. Sumpter.
\newblock {\em Collective animal behavior}.
\newblock Princeton University Press, 2010.

\bibitem{herbert-read.j:2015}
J.E. Herbert-Read, M.~Romenskyy, and D.J.T. Sumpter.
\newblock A turing test for collective motion.
\newblock {\em Biology Letters}, 11:20150674, 2015.

\bibitem{lobaskin.v:2013}
V.~Lobaskin and M.~Romenskyy.
\newblock Collective dynamics in systems of active brownian particles with
  dissipative interactions.
\newblock {\em Phys. Rev. E}, 87:052135, 2013.

\bibitem{grossmann12}
R~Gro{\ss}mann, L~Schimansky-Geier, and P~Romanczuk.
\newblock Active brownian particles with velocity-alignment and active
  fluctuations.
\newblock {\em New Journal of Physics}, 14(7):073033, 2012.

\bibitem{Katz198120}
Lawrence~C. Katz, Michael~J. Potel, and Richard~J. Wassersug.
\newblock Structure and mechanisms of schooling intadpoles of the clawed frog,
  xenopus laevis.
\newblock {\em Animal Behaviour}, 29(1):20 -- 33, 1981.

\bibitem{OBRIEN19891}
D.P. O'Brien.
\newblock Analysis of the internal arrangement of individuals within crustacean
  aggregations (euphausiacea, mysidacea).
\newblock {\em Journal of Experimental Marine Biology and Ecology}, 128(1):1 --
  30, 1989.

\end{thebibliography}

\begin{thebibliography}{4}
\expandafter\ifx\csname natexlab\endcsname\relax\def\natexlab#1{#1}\fi
\expandafter\ifx\csname bibnamefont\endcsname\relax
  \def\bibnamefont#1{#1}\fi
\expandafter\ifx\csname bibfnamefont\endcsname\relax
  \def\bibfnamefont#1{#1}\fi
\expandafter\ifx\csname citenamefont\endcsname\relax
  \def\citenamefont#1{#1}\fi
\expandafter\ifx\csname url\endcsname\relax
  \def\url#1{\texttt{#1}}\fi
\expandafter\ifx\csname urlprefix\endcsname\relax\def\urlprefix{URL }\fi
\providecommand{\bibinfo}[2]{#2}
\providecommand{\eprint}[2][]{\url{#2}}

\bibitem[{\citenamefont{Handegard and Williams}(2008)}]{handegard.n:2008}
\bibinfo{author}{\bibfnamefont{N.~O.} \bibnamefont{Handegard}}
  \bibnamefont{and} \bibinfo{author}{\bibfnamefont{K.}~\bibnamefont{Williams}},
  \bibinfo{journal}{ICES J. Marine Sci.: Journal du Conseil}
  \textbf{\bibinfo{volume}{65}}, \bibinfo{pages}{636} (\bibinfo{year}{2008}),
  \urlprefix\url{http://icesjms.oxfordjournals.org/content/65/4/636.abstract}.

\bibitem[{\citenamefont{Weisstein}()}]{weinstein.ew}
\bibinfo{author}{\bibfnamefont{E.~W.} \bibnamefont{Weisstein}},
  \emph{\bibinfo{title}{Circle-circle intersection}},
  \bibinfo{howpublished}{MathWorld -- A Wolfram Web Resource.
  \url{http://mathworld.wolfram.com/Circle-CircleIntersection.html}}.

\bibitem[{\citenamefont{Barber et~al.}(1996)\citenamefont{Barber, Dobkin, and
  Huhdanpaa}}]{Barber:1996}
\bibinfo{author}{\bibfnamefont{C.~B.} \bibnamefont{Barber}},
  \bibinfo{author}{\bibfnamefont{D.~P.} \bibnamefont{Dobkin}},
  \bibnamefont{and}
  \bibinfo{author}{\bibfnamefont{H.}~\bibnamefont{Huhdanpaa}},
  \bibinfo{journal}{ACM Trans. Math. Softw.} \textbf{\bibinfo{volume}{22}},
  \bibinfo{pages}{469} (\bibinfo{year}{1996}), ISSN \bibinfo{issn}{0098-3500},
  \urlprefix\url{http://doi.acm.org/10.1145/235815.235821}.

\bibitem[{\citenamefont{Bourke}()}]{bourke.p}
\bibinfo{author}{\bibfnamefont{P.}~\bibnamefont{Bourke}},
  \emph{\bibinfo{title}{Calculating the area and centroid of a polygon}},
  \bibinfo{howpublished}{\url{http://paulbourke.net/geometry/polygonmesh/}}.

\end{thebibliography}
\end{document}